# Novel electronic behavior facilitating the NdNiO₃ metal-insulator transition

M. H. Upton[*1], Yongseong Choi[1], Hyowon Park[2], Jian Liu[3,4], D. Meyers[5], S. Middey[5], Jong-Woo Kim[1], Philip J. Ryan[*1]

[1]Advanced Photon Source, Argonne National Laboratory, Argonne, Illinois 60439, USA
[2] Department of Physics, University of Illinois at Chicago 60607, USA
[3] Advanced Light Source, Lawrence Berkeley National Laboratory, Berkeley, California 94720, USA
[4] Materials Science Division, Lawrence Berkeley National Laboratory, Berkeley, California 94720, USA
[5] Department of Physics, University of Arkansas, Fayetteville, Arkansas 72701, USA

**We present evidence that the metal-insulator transition (MIT) in a tensile strained NdNiO₃ (NNO) film is facilitated by a redistribution of electronic density and neither requires Ni charge disproportionation nor symmetry change [1, 2]. Given epitaxial tensile strain in thin NNO films induces preferential occupancy of the $e_g$ $d_{x^2-y^2}$ orbital ($d_{3z^2-r^2}$) we propose the larger transfer integral of this orbital state with the O 2p mediates a redistribution of electronic density from the Ni atom. A decrease in Ni $d_{x^2-y^2}$ orbital occupation is directly observed by resonant inelastic x-ray scattering below the MIT temperature. Furthermore, an increase in Nd charge occupancy is measured by x-ray absorption at the Nd L₃ edge. Both spin-orbit coupling and crystal field effects combine to break the degeneracy of the Nd 5d states shifting the energy of the Nd $e_g$ $d_{x^2-y^2}$ orbit towards the Fermi level allowing the A site to become an active acceptor during the MI transition. This work identifies the relocation of electrons from the Ni 3d to the Nd 5d orbitals across the MIT. We propose the insulating gap opens between the Ni 3d and O 2p resulting from Ni 3d electron localization mediated by charge loss. The transition seems neither purely Mott-Hubbard nor simple charge transfer.**

**Introduction**

The discovery of superconductivity sparked interest in emergent phenomena in correlated electron materials, including metal insulator transitions [3]. Many different models including the Mott-Hubbard, Mott-Heisenberg, charge-transfer, covalent insulator and Slater insulator have been suggested to explain the phenomena of a MIT in a material with partially filled *d*-bands. The proximate cause of a transition in a particular material, however, can be difficult to determine due to the delicate balance of forces experienced by the electrons, including intra-electron, spin-spin, spin-orbital and electron-lattice interactions. Epitaxial growth offers the opportunity to test different MIT theories as strain modifies the interatomic distances of chemically identical compounds and can thus isolate interactions that drive the MIT [4].

Temperature driven MIT in NdNiO₃ (NNO) occurs at ~200K. Many different electronic and structural rearrangements accompany this transition. Anti-ferromagnetic order [5]; structural changes, including an increase of orthorhombic rotations in concert with a small volume increase [6]; Ni charge disproportionation and redistribution mediated by oxygen have been reported in bulk samples [1].

Interpretation of the MIT phenomenon centers upon whether the transition falls under a Mott-Hubbard [9, 10] or charge transfer paradigm [11]. In a Mott-Hubbard phase transition the repulsive Coulomb force between electrons (U) causes the electrons to localize opening a band gap between the occupied and unoccupied *d* electron states of the transition metal. In the charge transfer mechanism the band gap opens between the transition metal *d* states and the oxygen 2*p* orbitals. In addition Stewart et al. proposed that the insulating gap is not closed by evolving changes in the Ni or O states immediately above and below the gap, but rather by the emergence or activation of electronic states in the insulating gap [12]. Yamamoto and Fujiwara calculated that the details of the Ni-O bond with AFM ordering in bulk NNO find a gap opening [13], in reasonable agreement with the measured insulating gap [14].

This work demonstrates that charge order is not requisite for the MIT and eliminates models based upon this phenomenon. Additionally, we propose a subtle charge redistribution process occurs coincident with the MIT. In tensile strained thin NNO films grown on SrTiO₃ (001) we found neither, evidence of charge disproportionation, symmetry change, nor an increase of octahedral rotations across the MIT transition. However a charge-relocation from the Ni to the Nd atom is seemingly observed across the transition. We propose a partial B-site charge transfer to the A-site *5d* state.

[*]Corresponding authors mhupton@aps.anl.gov and pryan@aps.anl.gov.

Furthermore, the migration of charge from the Ni in the insulating state combined with a lattice expansion suggests the remaining Ni *d* electrons will be increasingly localized, contributing to the opening of an insulating gap.

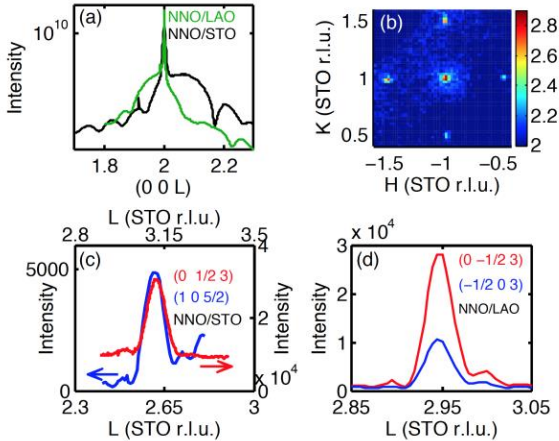

*Figure 1 a)* L-scan through the (002) reflections for both films. b) Reciprocal space map around the (113) reflection showing four orthorhombic half integer reflections, (1 ½ 3), (½ 1 3), etc. c) & d) Multiple orthorhombic domains observed in both samples.

**Experimental Results.**

Recent transport measurements demonstrated that tensile strain in NNO preserves the MIT, while compressive strain extinguishes the insulating phase [15]. Two epitaxial NNO films, prepared by PLD [15], are studied in this work. The first is 10 unit cells grown on $LaAlO_3$ (001) (LAO) in an almost unstrained state (0.3% compressive), shows no MIT or temperature dependent structural change. The second sample is a fully strained, 30 unit cell film grown on $SrTiO_3$ (001) (STO), in a tensile strain state (2.6%). The NNO/STO MIT temperature is reduced from the bulk value to ~180 K [15]. Both films are fully strained to their respective substrates while the out-of-plane lattice parameters are 3.84 Å and 3.75 Å on LAO and STO respectively [15]. Structural changes in both films are measured with x-ray diffraction while electronic changes are measured with resonant x-ray diffraction, x-ray absorption spectroscopy (XAS) and resonant inelastic x-ray scattering (RIXS). All experiments were performed at the Advanced Photon Source, Argonne National Laboratory.

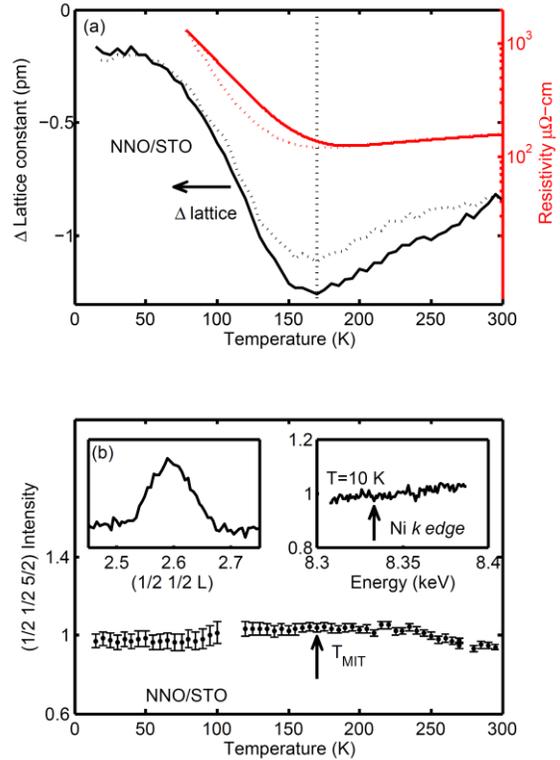

*Figure 2 a)* The left axis shows the response in the c-axis lattice constant of NNO/STO (NNO/LAO) to heating with a solid purple (green) line. The dotted purple/ green lines show the response to cooling. The dotted vertical line indicates the MIT temperature. The right axis shows the resistivity of NNO/STO as a function of temperature in response to heating (cooling) with a solid (dashed) line b) Temperature dependence of the (1/2 1/2 5/2) reflection. The upper left inset L scan through (1/2 1/2 5/2) peak (a combination of (1 0 5)o, (0 1 5)o, (2 3 1)o and (3 2 1)o for the orthorhombic notation from the twinned domains). Top right inset shows no Ni K response at this reflection.

Specular rod (L) scans through (0 0 2) reflections from both samples are shown in Fig. 1a. In the tensile strained film, the orthorhombic c-axis lies along all three a, b and c crystallographic directions of the STO substrate, while, for the LAO substrate, it is confined to the substrate plane (Fig. 1b, d). In NNO/LAO the out-of-plane lattice constant of the film undergoes thermal contraction with decreasing temperature and shows no transitional expansion. In the case of NNO/STO the out-of-plane lattice constant of the film increases coincident with the MIT. Bulk NNO also exhibits a volume expansion (primarily derived from the orthorhombic b direction), similar in magnitude to what is observed in the NNO/STO film [6]. The (1/2 1/2 5/2) diffraction peak from NNO/STO, which constitutes primarily octahedral rotations, is presented in Fig. 2b. There is no evidence of a significant intensity change as the sample passes through the MIT or as the incident energy is tuned through the Ni K-edge. This indicates that, unlike bulk NNO, no charge order emerges below the transition temperature. Subsequent measurements on single domain NNO found the same result [16]. Previous Ni K resonant diffraction interpreted their results as evidence of emergent charge order resulting from a Ni charge disproportionation and posited that it drove the MIT [1]. Density functional calculations have however suggested that such charge disproportionation is not necessary for the MIT in LuNiO$_3$ [9].

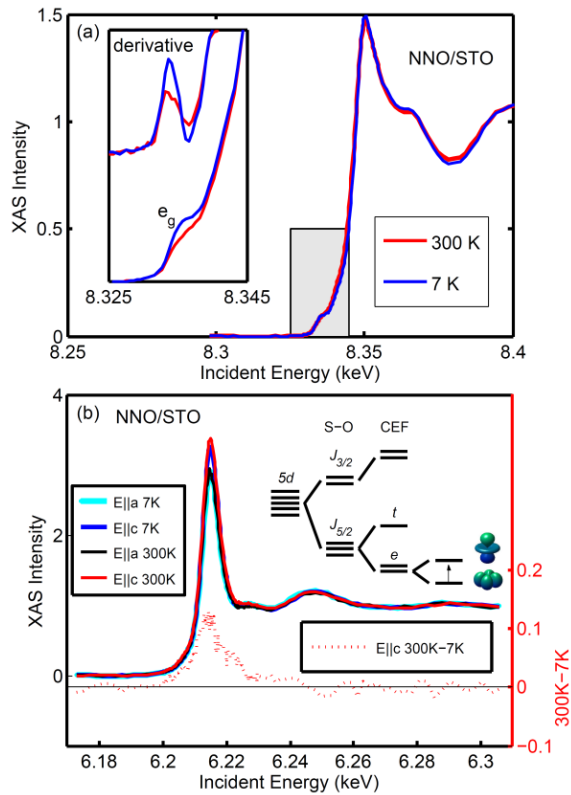

Figure 3a) XAS measurements at the Ni K edge show a clear increase in pre-edge intensity in the low temperature phase indicating an increase in unoccupied $e_g$ states in the insulating phase. b) XAS measurements at the Nd L$_3$-edge show polarization dependence as a result of the split and singly occupied Nd d states. The XAS spectra show a greater Nd electron occupation below the MIT than above.

Temperature dependent XAS spectra at the Ni *K* edge from NNO/STO are shown in Fig. 3a. Below the transition temperature there is an increase in the intensity of the Ni K pre-edge peak. The pre-edge peak results from a *1s* to *3d* ($e_g$) transition. Nominally, the Ni electronic configuration is Ni 3$d^7$. It has been determined, however, that it is energetically favorable to transfer partial hole from the Ni to the hybridized Ni-O bond. Thus, the ground state is generally described as mixture of $|3d^72p^6\rangle$ and $|3d^8\underline{L}\rangle$, where $\underline{L}$ is the ligand hole [17, 18, 19]. In the electronic state $|3d^8\underline{L}\rangle$ a hole is donated from the O *2p* states to the Ni *3d* states. The latter, however, may be hybridized with the O *2p* and thus can be better understood as Ni *3d* with some O *2p* character [20]. The $\underline{L}$ consists of $p\sigma_z$ or $p\sigma_{x,y}$ orbitals [21]. An increase in the pre-edge peak intensity may be interpreted as a decrease in the Ni $e_g$ occupation. Thus the metallic phase has greater Ni $e_g$ occupation than the insulating phase. Naively it might be assumed that the electrons are exchanged with the O ions. XAS measurements of NNO at the O edge, however, do not show any measurable change through the transition temperature [18], suggesting that this is not the case.

To complete the picture of the electronic configuration, XAS measurements at the Nd L$_3$ edge were performed, Fig. 3b. A significant response of the white line was measured from the NNO/STO sample as a function of incident x-ray polarization. The degenerate *5d* states of the Nd ion are split by a combination of spin-orbit coupling and crystal field, as illustrated in the inset. For rare earths, the strong spin-orbit coupling from the *4f* orbital influences the *5d* states via the *4f-5d* exchange interaction [22]. As a result, the J$_{5/2}$ state is driven towards the Fermi energy, becoming potentially active. Therefore, at the L$_3$ edge, the XAS spectra show polarization dependence for the incident x-ray while no difference is observed at the L$_2$ edge (not shown).

The XAS measurements at the Nd L$_3$ edge also reveal a small change above and below the transition temperature. This is unexpected because the A site is not, in the literature, typically ascribed a role in the MIT. When cooled below the transition temperature, the white line intensity decreases in the out-of-plane (E//c) measurement, which is indicative of an increase of the electron occupation for the Nd *5d* state. Combined with the observed reduction of the Ni *3d* occupation these results suggest a partial electron transfer from the Ni *3d* to the Nd *5d* across the MIT.

Electron transfer from Ni to Nd naively seems unlikely because a large energy separation is typically calculated between A and B site electrons in transition metal oxides. Experimentally, however, there is clear evidence of Ni-Nd interaction. Femtosecond resonant soft x-ray diffraction with high-resolution spectroscopic analysis has shown the low-temperature magnetic dynamics of the Ni and Nd lattices are coupled [23]. Indeed, a second set of resonant soft x-ray diffraction measurements have shown that the Nd magnetic order is induced by the Ni magnetic order [24]. Furthermore, in the case of another rare-earth-*3d* perovskite material, EuTiO₃, direct A-B site hybridization has been measured [25] and calculated [26].

To explore the microscopic origins of Ni electron behavior resonant inelastic x-ray spectroscopy (RIXS) measurements are performed at the Ni K pre-edge. RIXS spectra from NNO/STO at a variety of incident energies are shown in Fig. 4a. The energy loss of the peaks remains constant regardless of incident energy, indicating a true RIXS signal and the measurement at the pre-edge incident energies indicates a direct, as opposed to indirect, RIXS process at work.

While both low spin [18, 27, 6] and high spin [9] configurations have been presented in the literature this data is most consistent with a low spin configuration with an *e_g* occupation less than 2 (>1). The experiments were performed at sector 9 at the APS using a spherical (1m radius) diced Ge(6 4 2) analyzer is used to obtain an overall energy resolution of 200 meV. To minimize the elastic background, most of the measurements were carried out in a horizontal scattering geometry. The Ni K pre-edge excites a *1s* core electron to an empty *3d* state, the excited system quickly decays: a d electron from below the Fermi level fills the *1s* core hole and a photon is emitted. The energy difference between the emitted and incident photon is equivalent to the energy difference between an unoccupied and occupied d state, a dd excitation [28, 29].

The NNO/LAO RIXS spectrum, shown in Fig. 4d, shows one dd excitation, at 0.79 eV energy loss, which is identified as the energy difference between *e_g* and *t_2g* electron energies and is thus a measure of the crystal field splitting [12]. The RIXS spectrum from NNO/STO presents two peaks at 0.6 eV and 1.0 eV (Fig. 3d). The emergence of two dd transitions is attributed to strain-induced *e_g* splitting while the one peak is observed in the 0.3% compressively strained film, where the expected *e_g* splitting is beyond resolution. Strain imposes a tetragonal distortion on the unit cell, resulting in a modification of the Ni-O bond lengths in the NiO₆ octahedra, as shown in the inset of Fig. 4c. This bond distortion reduces the crystal field symmetry and removes the two-fold *e_g* orbital degeneracy. In the case of tensile strain the in-plane lattice expands, reducing the energy of the $d_{x^2-y^2}$ orbital and conversely the out-of-plane $d_{3z^2-r^2}$ orbital rises with the reduced c-axis lattice parameter.

The loss energies do not change through the volume expansion of the MIT indicating that the relative energies of the orbitals are constant within the resolution of the experiment. The lower energy intensity ($d_{x^2-y^2}$) however, increases in the insulating phase while the higher energy ($d_{3z^2-r^2}$) intensity remains unchanged (as illustrated in Fig. 3e). The former effect ($d_{x^2-y^2}$) indicates an increased probability of a Ni *1s* to $d_{x^2-y^2}$ transition, and therefore a decrease in the ground state occupancy of the $d_{x^2-y^2}$ orbital.

Discussion

This work illustrates that the NNO MIT occurs without concurrent Ni charge ordering. Additionally, we observe a redistribution of charge from the Ni *3d* to the Nd *5d* states across the MIT. The redistribution of electronic density from the Ni $d_{x^2-y^2}$ orbital occurs in films under tensile strain. In contrast, compressive strain shifts the $d_{3z^2-r^2}$ energy lower, making it the preferentially occupied *e_g* orbital. However, the $d_{3z^2-r^2}$ orbital has less d-pσ overlap, and thus less oxygen hybridization. Consequently the compressive state is unlikely to present similar electron relocation. Because the MIT electron transfer mechanism is associated with films under tensile strain the occupation of the Ni $d_{x^2-y^2}$ orbital and its' greater hybridization with the O *2p* orbitals may underlie the MIT mechanism.

This work reveals a process which may assist the insulating gap to open, as a result of Ni *3d* localization. Two effects likely contribute to increased Ni *3d* localization: the decrease in Ni electron occupation and the increase in Ni-O distance. The novel involvement of the A-site cation as an acceptor facilitates the charge transfer process resulting in Ni charge redistribution. Thus, the MIT is neither pure Mott-Hubbard, despite electron localization, nor simple charge-transfer. In addition, our results may aid in describing the proposed emergent mid gap state activity [12] through a partial transfer from Nd *5d* states to the Ni-O hybridized conduction band in conjunction with the metallic phase onset.

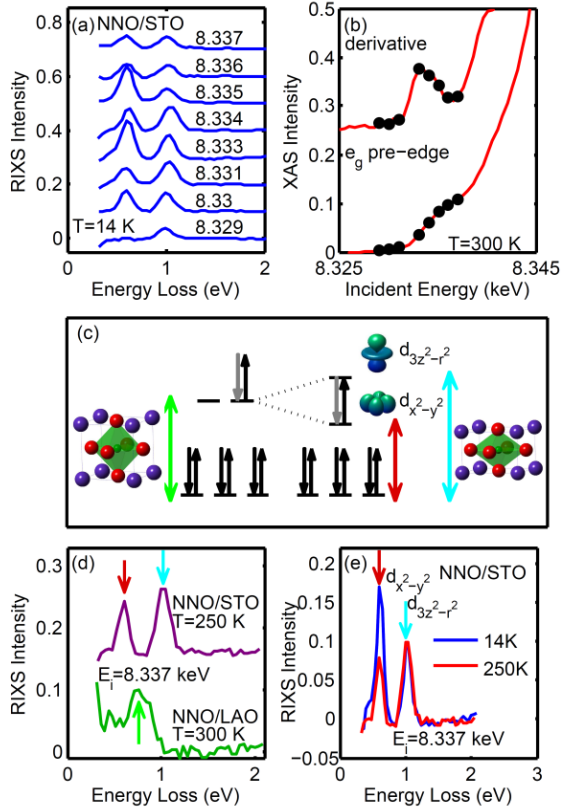

Figure 4a) RIXS spectra from NNO/STO measured at a variety of incident energies. b) Incident RIXS energies. c) A sketch showing the compression induced splitting of the Ni $e_g$ orbitals. One electron, partially donated by the O 2p, is shown in gray instead of black. d) RIXS spectra of NNO/STO and NNO/LAO measured at the Ni K pre-edge. The NNO/LAO spectra has degenerate $e_g$ levels, and therefore one dd transition between $t_{2g}$ and $e_g$ levels. The NNO/STO $e_g$ levels are split by tensile strain and therefore two dd transitions are observed. e) RIXS spectra of NNO/STO measured above and below the MIT. The low energy peak has higher intensity in the insulating phase as electron density from the $d_{x^2-y^2}$ orbital shifts away from Ni to Nd.

**Acknowledgements.**

Work at the Advanced Photon Source is supported by the US Department of Energy, Office of Science under Grant No. DEAC02-06CH11357. D. Meyers and S. Middey were supported by DOD-ARO under the Grant No. 0402-17291.